# Possible signatures of mixed-parity superconductivity in doped, polar SrTiO$_3$ films

**Timo Schumann\*, Luca Galletti\*, Hanbyeol Jeong, Kaveh Ahadi, William M. Strickland, Salva Salmani-Rezaie, and Susanne Stemmer[1)]**

Materials Department, University of California, Santa Barbara, California 93106-5050, USA

\* These authors contributed equally.

[1)] Corresponding author. Email: stemmer@mrl.ucsb.edu

**ABSTRACT**

Superconductors that possess both broken spatial inversion symmetry and spin-orbit interactions exhibit a mix of spin singlet and triplet pairing. Here, we report on measurements of the superconducting properties of electron-doped, strained $SrTiO_3$ films. These films have an enhanced superconducting transition temperature and were previously shown to undergo a transition to a polar phase prior to becoming superconducting. We show that some films show signatures of an unusual superconducting state, such as an in-plane critical field that is higher than both the paramagnetic and orbital pair breaking limits. Moreover, nonreciprocal transport, which reflects the ratio of odd vs. even pairing interaction, is observed. Together, these characteristics indicate that these films provide a tunable platform for investigations of unconventional superconductivity.

Topological superconductors can host quasiparticle states that offer a promising route for generating quantum entangled states that are protected against decoherence [1]. While many of the approaches that are currently being pursued involve hybrid structures [2-5], the discovery of an *intrinsic* topological superconductor would be an exciting step forward. One route involves finding superconductors that possess both broken spatial inversion symmetry and spin-orbit interaction, which are expected to exhibit a mix of singlet and triplet pairing due to the lifting of the two-fold spin degeneracy [6-9]. The application of a Zeeman field or other microscopic interactions can suppress the s-wave channel in favor of odd-parity superconductivity [8].

Bulk, unstrained $SrTiO_3$ is a multiband superconductor [10-12] that exhibits a superconducting dome with a maximum transition temperature ($T_C$) of ~ 300 mK at a carrier density of $1\times10^{20}$ cm$^{-3}$ [13]. The degeneracy of the three $t_{2g}$-derived conduction bands at the $\Gamma$ point is lifted by the low temperature tetragonal distortion and by spin-orbit coupling [14, 15]. With increasing carrier density, the three bands fill consecutively [15-17]. Quasi-two-dimensional electron systems at interfaces involving $SrTiO_3$, which exhibit signatures of Rashba spin-orbit coupling [18-22], have been suggested as possible candidates for topological superconductivity [8, 23]. Bulk $SrTiO_3$ is also a quantum paraelectric [24] that is near a ferroelectric transition. The fixed polarization charge of ferroelectrics, and corresponding electric fields, can easily rival those of high-density two-dimensional electron gases [25]. Early theoretical work already pointed out that ferroelectric order would be a promising route to realize odd-parity superconductivity [6]. Combining the ingredients of multiorbital effects, ferroelectricity, and spin orbit coupling, a topological superconducting state has been predicted [26, 27].

Recently, it was shown that ferroelectric (polar), doped SrTiO$_3$ can become superconducting [28-32]. In particular, doped SrTiO$_3$ films grown on (001) (LaAlO$_3$)$_{0.3}$(Sr$_2$AlTaO$_6$)$_{0.7}$ (LSAT), which are compressively strained, undergo a transition to a ferroelectric phase prior to becoming superconducting and show enhanced superconducting transition temperatures [28, 29]. Because of the broken inversion symmetry, such films are promising candidates in the search for an intrinsic topological superconductor.

A key signature of odd or mixed parity superconductors is an enhancement of the upper critical field beyond the paramagnetic (Clogston-Chandrasekhar or Pauli) limit [8]. Here, we show that some polar, superconducting SrTiO$_3$ films show anisotropic critical fields above the Pauli limit. Moreover, we find pronounced non-reciprocal charge transport, a signature of a noncentrosymmetric, spin-orbit coupled superconductor that is sensitive to the parity of the superconducting order parameter [33, 34]. Taken together, these observations indicate a highly unconventional superconductor.

Doped SrTiO$_3$ films were grown by hybrid metal-organic molecular beam epitaxy [35, 36]. Strained films with thicknesses between 160 nm and 180 nm were grown on (001) (LaAlO$_3$)$_{0.3}$(Sr$_2$AlTaO$_6$)$_{0.7}$ (LSAT) crystals. The lattice mismatch between SrTiO$_3$ ($a$ = 3.905 Å) and LSAT ($a$ = 3.868 Å) is -0.95 %. Some of the films were coherently strained (such as film A discussed below), while others (such as the films denoted B and D in the following) were partially relaxed, as discussed below. Details of the films' structure can be found in the Supplementary Information [37]. All films were $n$-type doped by substituting Sr$^{2+}$ with either La$^{3+}$ or Sm$^{3+}$. The choice of dopant had no influence on the superconducting parameters, such as critical field ($H_{c2}$) and $T_C$. Electrical measurements were performed on Hall bar devices with dimensions of 100 μm × 100 μm, which were fabricated by standard optical photolithography,

Ar ion milling for mesa definition and electron beam deposition of ohmic Ti/Au contacts. Hall measurements to extract carrier concentrations were performed between 300 K and 2 K in Quantum Design Dynacool systems. Here, $n_{3D}$ refers to volume carrier concentration estimated from the Hall measurements at 300 K and the film thickness. Hall bars were aligned along <100> and <010>, respectively, to facilitate measurements with **I** ∥ **H** and **I** ⊥ **H** with **H** in the film plane, where **H** is the magnetic field and **I** is the excitation current.

Measurements between 1 K and 12 mK were performed in an Oxford Instruments Triton dilution refrigerator using low-frequency lock-in techniques. For measurements of the first harmonic resistances, cryogenic filters were used on the current and voltage lines to reduce the temperature of the electron bath (see Supplementary Information [37]). Superconducting transitions were recorded on different devices and/or contact configurations for all samples to ensure reproducibility.

A nonlinear voltage response to an applied current is a manifestation of nonreciprocal currents in noncentrosymmetric Rashba superconductors [34, 38]. The voltage (*V*) response is given as $V = RI + \gamma RBI^2$ for **I** ⊥ **H**, where *R* is the resistance and *B* is the magnetic flux density [39]. In contrast, *V* = 0 for **I** ∥ **H** [39]. Here, $\gamma$ is a coefficient that describes the strength of the magneto-chiral anisotropy [39, 40]. Given an applied AC current $I = \sqrt{2}I_0 \sin \omega t$, where $\omega$ and *t* are angular frequency and time, respectively, it follows that:

$$V = \sqrt{2}RI_0 \sin \omega t - \gamma RBI_0^2 \cos 2\omega t + \gamma RBI_0^2. \qquad (1)$$

Nonreciprocal currents were measured by detecting the amplitudes of first and second harmonic resistances with a lock-in amplifier. Following the Eq. (1), the first and second harmonic resistances are defined as $R^\omega = R$ and $R^{2\omega} = \gamma BRI_0\sqrt{2}$, respectively. The cut-off frequency of the cryogenic filters is on the order of a few tens of Hz, so they were not used for measurements of the second harmonics. For these measurements, lock-in amplifier measurements at a higher frequency were used to enhance the signal-to-noise ratio [37].

We investigated films with a range of carrier densities [see Fig. 1(a)]. Figure 1 shows $T_C$, $H_{c2}$ and their ratio as a function of $n_{3D}$ with **H** oriented either in or out of the film plane, respectively. Here, $T_C$ is the temperature for which the resistance reaches 5 % of normal state resistance and $H_{c2}$ is defined as the magnetic field value when the resistance reaches 95% of the normal state resistance. The sharp drop in $T_C$ at lower concentrations is due to effects from surface depletion [41].

The $T_C$ values [Fig. 1(a)] are in agreement with values previously reported for strained, doped films on LSAT [28]. Enhanced $T_C$ values (up to 600 mK) near the peak of the superconducting dome relative to unstrained, doped $SrTiO_3$ are directly connected to the ferroelectric normal state, as shown elsewhere [28, 29]. The films are polar (point group 4*mm*) with the electric polarization vector pointing normal to the film plane [29]. Partially relaxed films, such as the films B and D (see Supplementary Information [37]), show a slightly reduced $T_C$, which is, however, still larger than that of unstrained, doped $SrTiO_3$. In the following we will focus on slightly underdoped films with $n_{3D}$ tuned between $5.5 \times 10^{19}$ cm$^{-3}$ and $8 \times 10^{19}$ cm$^{-3}$.

Figure 1(b) shows $H_{c2}$ as a function of $n_{3D}$. In general, $H_{c2}$ increases with $n_{3D}$, as expected for films on the underdoped side of the dome. When **H** is in-plane, $H_{c2}$ is strongly enhanced; some samples only enter the normal conducting state at $\mu_0\mathbf{H} > 2$ T ($\mu_0$ is the vacuum permeability). These $H_{c2}$ values are comparable to the largest values previously reported for doped $SrTiO_3$ [21]. Figure 1 also shows the values of an unusual sample (film D) with an even more strongly enhanced in-plane $H_{c2}$. Film D only enters the normal conducting state at $\mu_0\mathbf{H} >$ 3.5 T. The main difference between films B and D is the presence of a thin layer near the interface with the substrate (see Supplementary Information [37]). It is likely that the very large in-plane $H_{c2}$ of film D is associated with the presence of this layer.

The ratio of $H_{c2}$ over $T_C$ is shown in Fig. 1(c). The dotted line marks the Clogston-Chandrasekhar (Pauli) limit, at which the condensation energy of the Cooper pairs, $\Delta = 1.78 k_B T_c$, equals the magnetic polarization energy of a spin singlet cooper pair $E = \frac{g \mu_B}{2} H_c$, where $\mu_B$ is the Bohr magneton, $g\sim2$ is the Landé factor and $k_B$ is Boltzmann's constant [42]. Expressed in units of $\mu_0 H/T$, this ratio is 1.85 T/K. For measurements with **H** normal to the film plane, the ratio is below the Pauli limit for most films, while for **H** in-plane, the ratio is above the limit for several films, and especially so for film D. While the breaking of the Pauli limit and a strong anisotropy of $H_{c2}$ is often taken as an indicator of spin triplet superconductivity [8], it should not be taken as the only evidence, especially in materials with strong spin orbit coupling [9], thin films [21, 43, 44], and multiband superconductors [11, 12].

The transition behavior of the normalized (to the normal state) resistance, $R/R_n$, is shown in Fig. 2 for the four films labelled A, B, C, and D in Fig. 1, measured with **H** in- and out-of-plane, respectively. Measurements with **H** perpendicular to the film plane (left column) show conventional behavior. In contrast, film D exhibits a "double-dome" structure for **H** in-plane. This shape drastically differs from the behavior described by the Werthamer-Helfand-Hohenberg (WHH) theory for singlet superconductors. Like the large in-plane $H_{c2}$, the double dome structure might be associated with distinct order parameters of the bulk and interface layer, respectively.

As an independent measurement supporting an unconventional superconducting state, we measured nonreciprocal transport. Figure 3(a) shows second harmonic resistances of film B. The second harmonic signal is purely imaginary, in agreement with Eq. (1). An antisymmetric component emerges in the second harmonic signal, only if $\mathbf{I} \perp \mathbf{H}$ (see Supplementary Information for a measurement with $\mathbf{I} \parallel \mathbf{H}$ [37]). The extracted $\gamma$ parameter for this sample is $(1.6\pm0.5)\times10^3$ $T^{-1}A^{-1}$. The error was calculated using error propagation and a conservative estimate of the errors in $R^{2\omega}$, $R^{\omega}$, and in determining the **H**-position of the peak. Figures 3(b,c) show the temperature dependence of the second harmonic signal and the extracted $\gamma$ parameter of sample D. The signal decreases in strength with increasing temperature and vanishes at temperatures above 300 mK. $R^{\omega}$ was recorded in the same measurement setup and used to the extract $\gamma$ parameter [Fig. 3(c)]. Here, $\gamma$ increases with decreasing temperature and reaches values of $(4.7\pm0.5)\times10^4$ $T^{-1}A^{-1}$, which is larger by a factor of six than values found in the noncentrosymmetric superconductor $MoS_2$ [39] and two orders of magnitude higher than that of an interfacial electron gas in $SrTiO_3$ [38]. The difference in $\gamma$ parameters between films B and D is mainly due to strongly enhanced $H_{c2}$ of film D. Thus, at the field where $R^{2\omega}$ peaks, which is similar for both films, $R^{\omega}$ of film D is smaller.

In spin-orbit coupled, noncentrosymmetric Rashba superconductors, $\gamma$ is proportional to the product of the strength of spin-orbit coupling and the ratio between the odd and even parity pairing [34]. Unlike films B and D, several other strained (and thus noncentrosymmetric) films, such as film A, showed no detectable nonreciprocal currents (see Supplementary Information [37]). A main result of this study is therefore that the lack of inversion symmetry alone is insufficient to produce a large nonreciprocal current and $\gamma$ value in polar, doped $SrTiO_3$ films whose normal-state ferroelectric polarization is normal to the film plane [29]. Additional microscopic parameters appear to play a role in determining the nature of the superconductivity. As discussed above, the superconducting properties of film D appear to have contributions from an interface region, resulting in the strongly enhanced in-plane $H_{c2}$ and $\gamma$ parameter. The nonreciprocal signal is, however, not solely an interface effect because it is also observed in film B, which does not have an interfacial layer. A common feature of both films B and D is partial strain relaxation [37]. Theoretical studies that clarify the role of interfaces, electronic structure, dimensionality, and strain gradients would be an important step in further understanding and tuning the superconducting properties of these films.

To briefly summarize, we find that some doped, polar $SrTiO_3$ films show signatures characteristic of spin-orbit coupled, noncentrosymmetric superconductors. Such a superconductor will exhibit a spin-triplet pairing component and thus has the potential to host non-Abelian Majorana bound states. Further development of existing models of topological superconductivity in $SrTiO_3$ [8, 23, 26, 27] to describe the polar films studied here would be very desirable. Ultimately, the topological nature of the superconducting phase(s) should be characterized using techniques that can discern that nature of edge states or bound states in the vortex cores.

**Acknowledgments**

The authors thank Tess Winkelhorst for help with the fabrication of the cryogenic filters. This work was supported by the U.S. Department of Energy (Award No. DE-SC0020305). The work made use of the MRL Shared Experimental Facilities, which are supported by the MRSEC Program of the US National Science Foundation under Award No. DMR 1720256.

**Figure Captions**

**Figure 1:** Superconducting properties as a function of $n_{3D}$ for different field orientations (IP = **H** in the film plane; OOP = **H** normal to the film plane). (a) $T_C$, (b) $H_{c2}$ and (c) the ratio of $\mu_0 H_{c2}$ over $T_C$. The dotted line in (c) indicates the Pauli limit. Arrows point to data from the four films (A, B, C, D) discussed in more detail. The open symbols are a film doped with Sm, all others are doped with La.

**Figure 2:** Normalized (to the normal state) longitudinal resistances, $R/R_n$, represented by the color scale, as a function of temperature ($T$) and applied magnetic field for films A, B, C, and D. The left columns show the data with **H** oriented perpendicular to the film plane, while the right columns are taken with **H** oriented in the film plane.

**Figure 3:** (a) Asymmetric component of the second order harmonic, $R^{2\omega}_{asym}$, for film B ($n_{3D}$ = $6.2 \times 10^{19}$ cm$^{-3}$). The magnetic field is in the sample plane and perpendicular to the applied current. The solid lines are obtained by smoothing over 50 data points. The antisymmetric component was obtained via $R^{2\omega}_{asym}(H) = (1/2)\bigl(R^{2\omega}(+|H|) - R^{2\omega}(-|H|)\bigr)$. (b) $R^{2\omega}_{asym}$ for film D as a function of temperature. Traces are offset by 30 mOhm for clarity. (c) Extracted values for the $\gamma$ parameter [see Eq. (1)] as a function of temperature for film D. Error bars were obtained via error propagation from standard deviations on the values of $R^{2\omega}$ and the position of the peak.

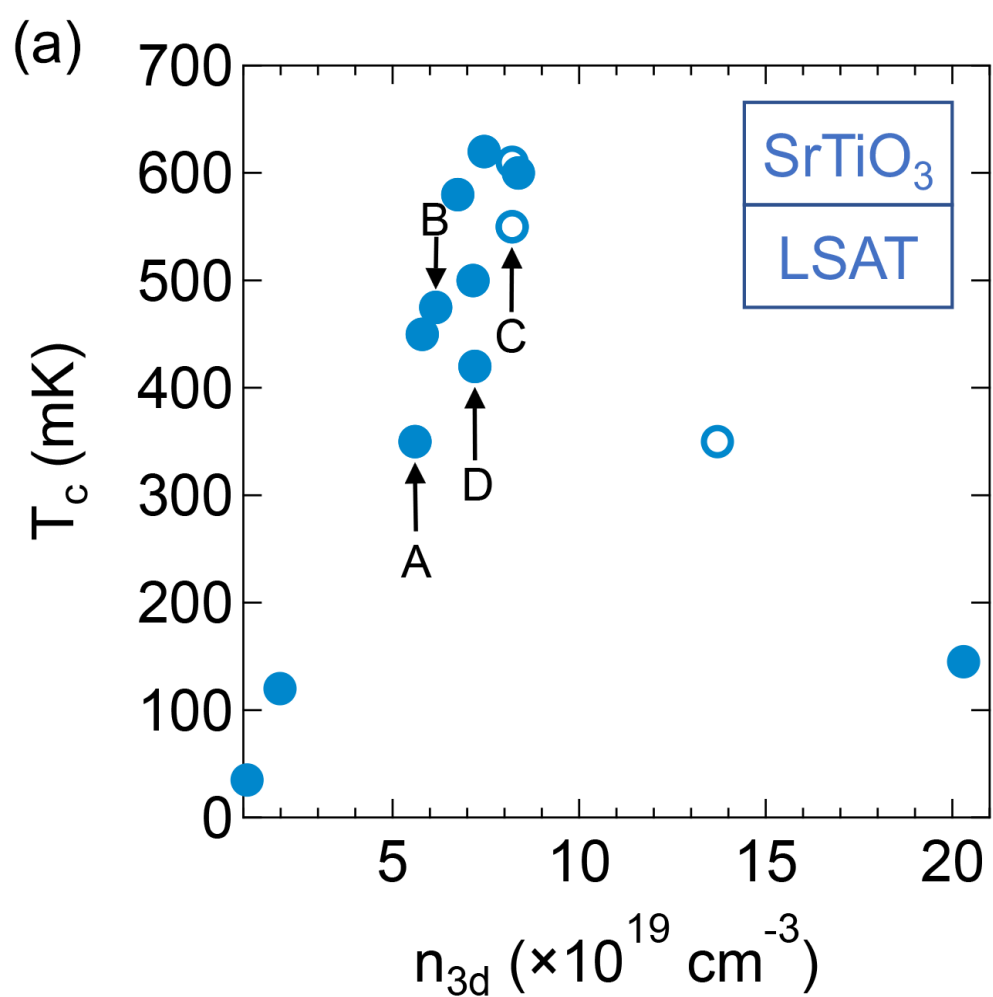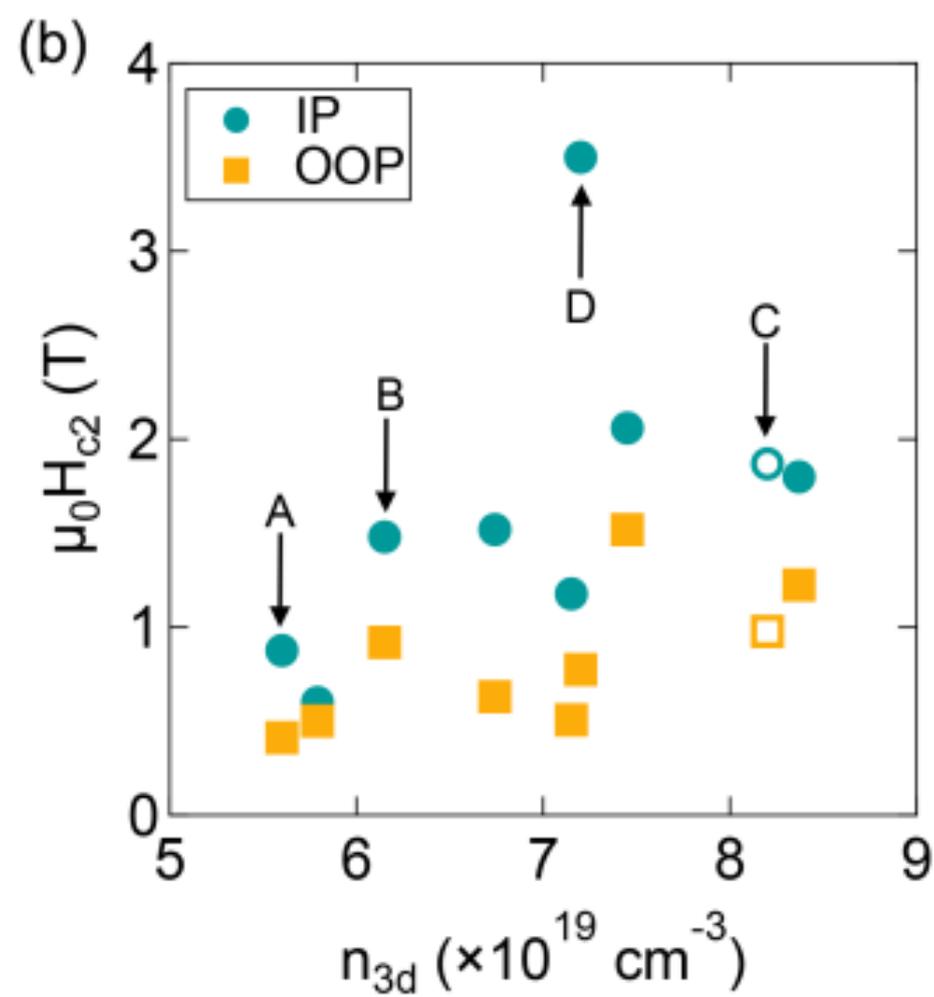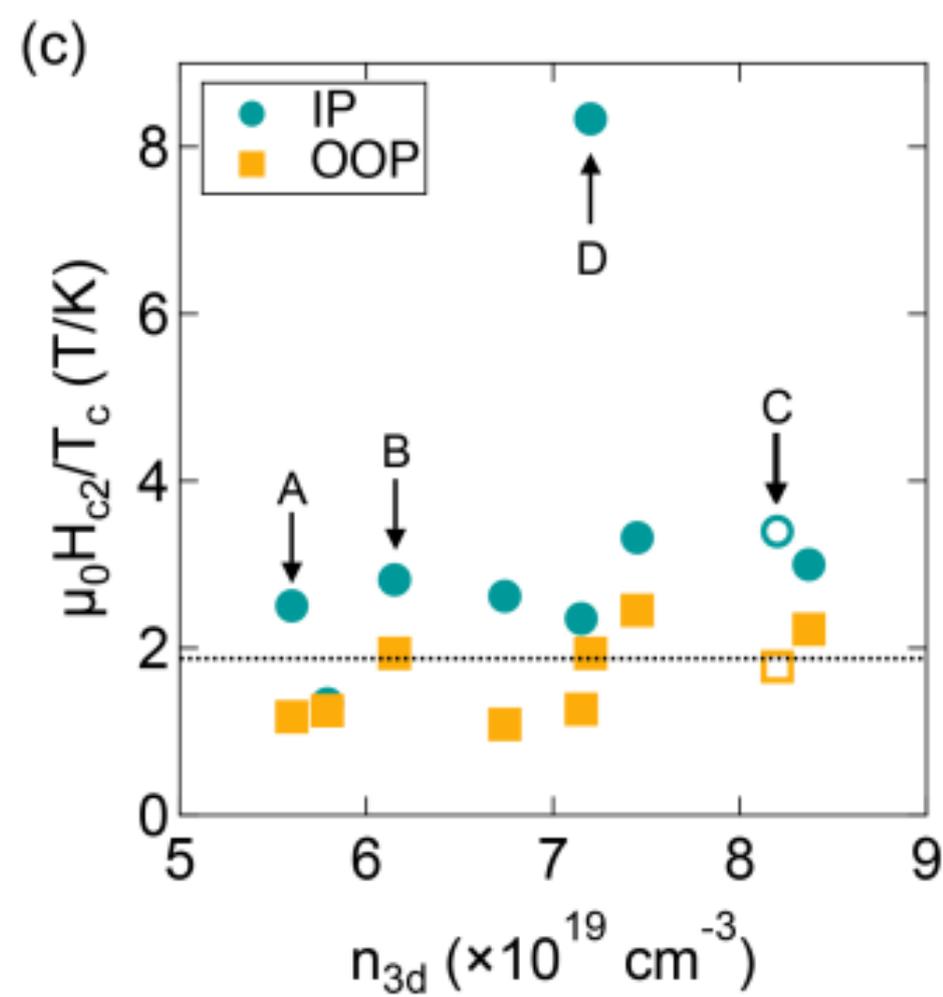

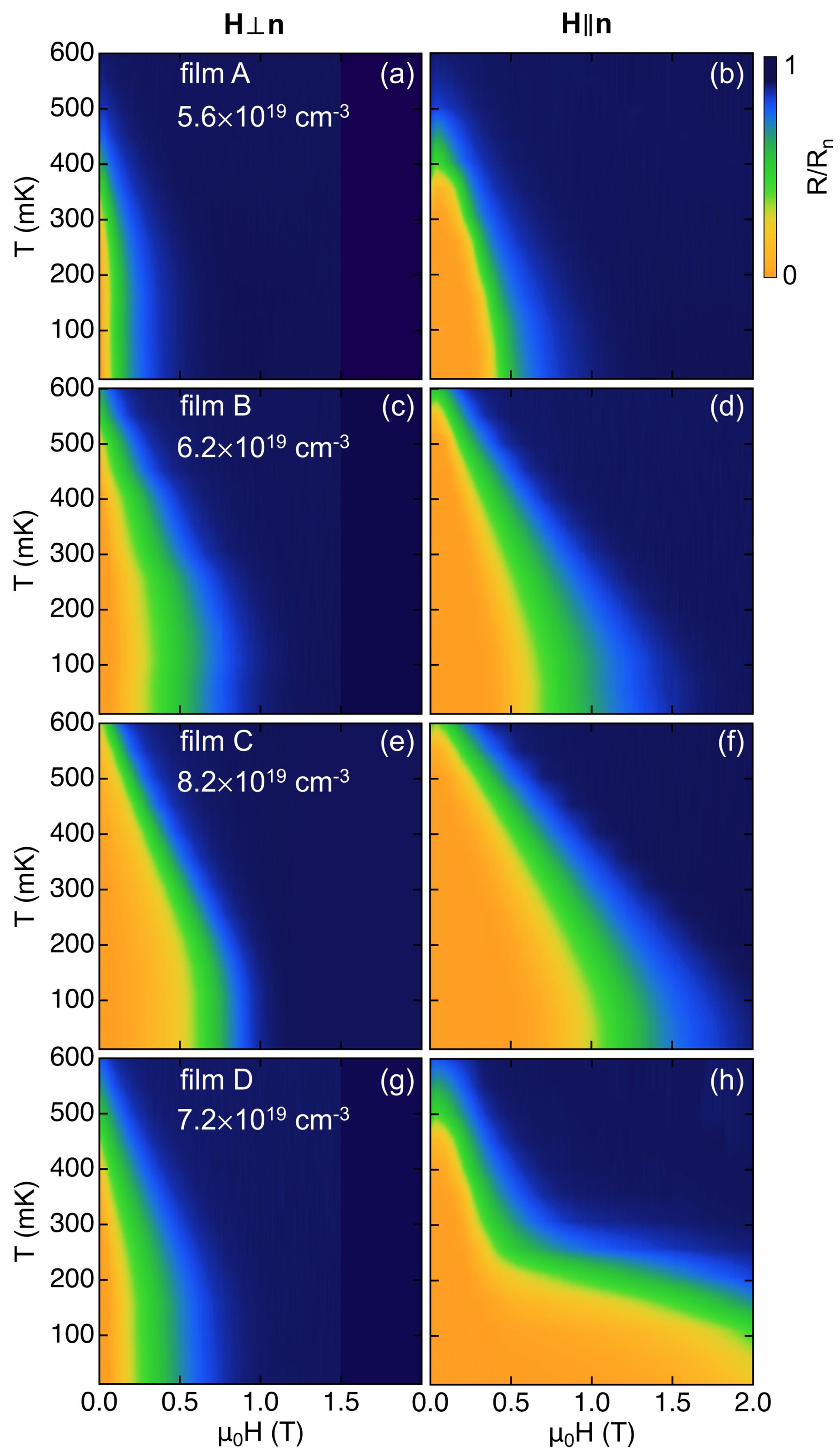

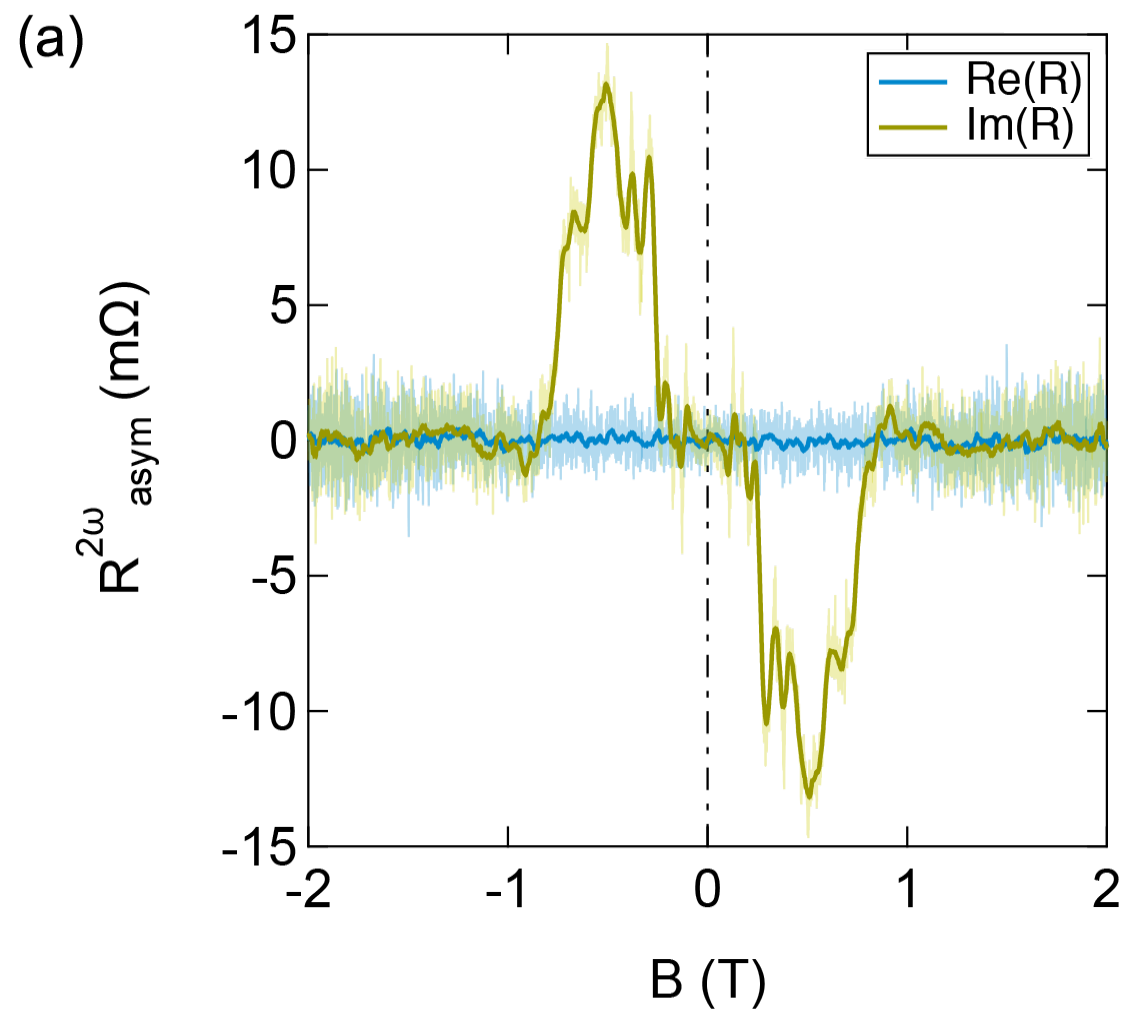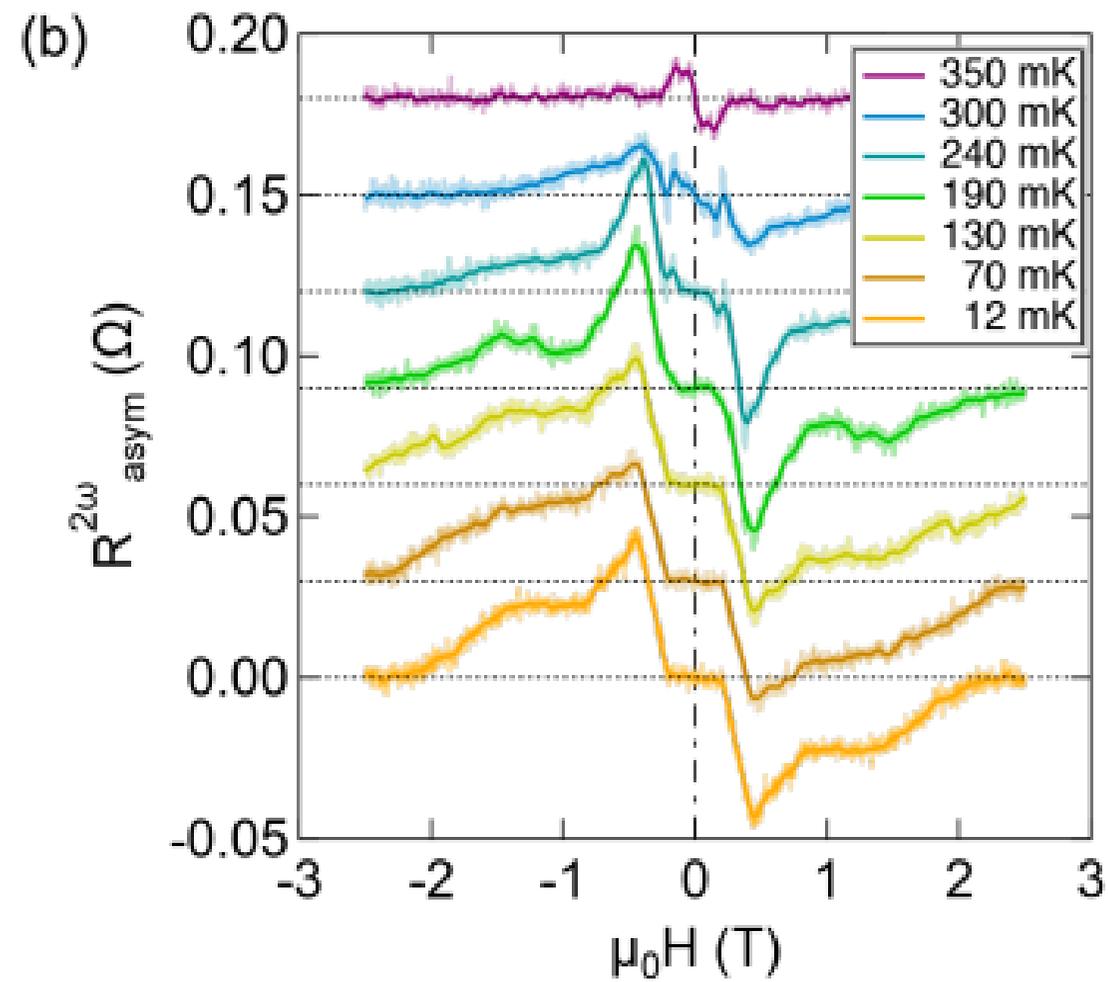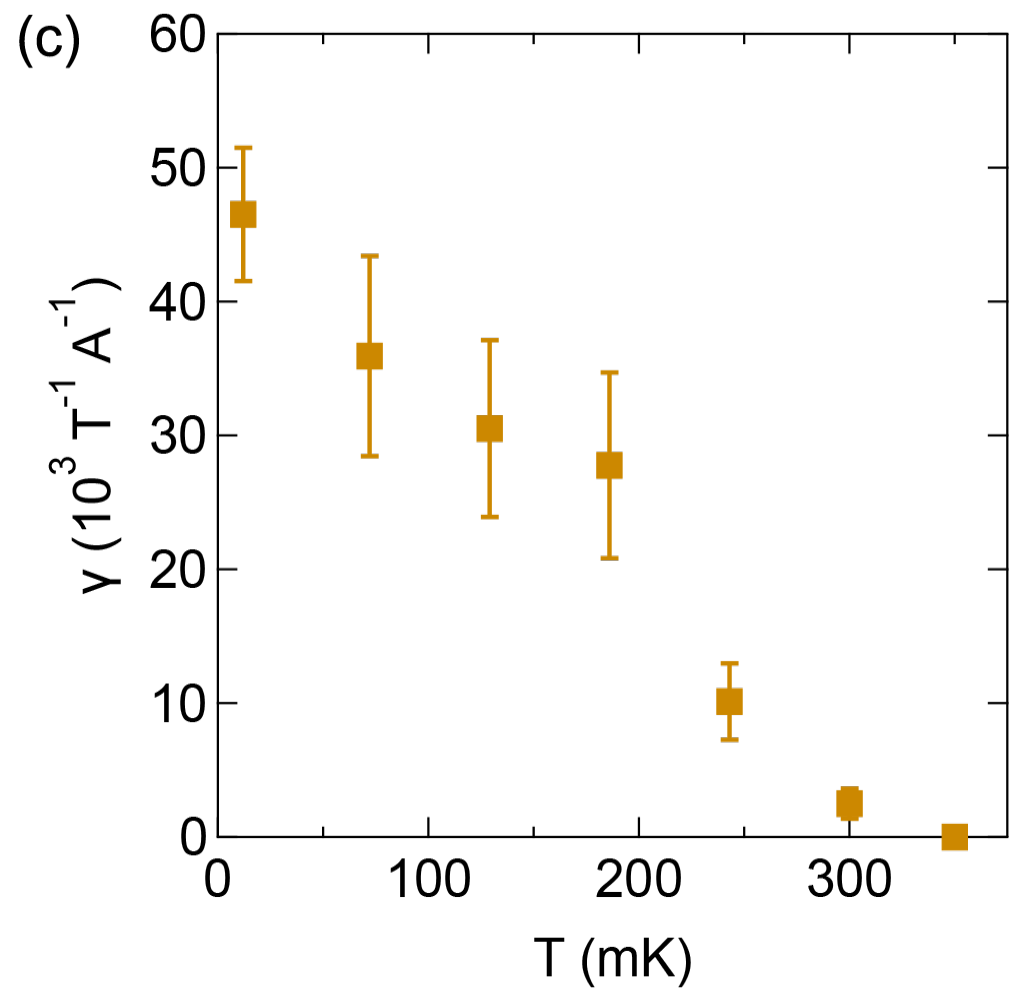